\documentstyle[epsfig,multicol,aps,prl]{revtex}

\begin{document}
\draft 
\begin{multicols}{2}

{\noindent \bf Dependence of the Fractional Quantum Hall Edge Critical Exponent on the Range of Interaction}
\vspace{0.3cm}

Recent experiments on external electron tunneling into an edge of a fractional quantum Hall (FQH) system find a striking current-voltage power law behavior $I\propto V^{\alpha}$, with exponent $\alpha\approx 2.7$ on the $f=\frac{1}{3}$ FQH plateau \cite{Chang96,Chang01}. Such power law $I$-$V$ characteristic with $\alpha =3$ was predicted for electron tunneling into an edge channel at the boundary of the $f=\frac{1}{3}$ FQH system: the low-energy dynamics is effectively 1D, and field-theoretic descriptions of edge channels as chiral Luttinger liquids have been developed \cite{Wen}. The equality $\alpha \!=\!2i+1$ for a FQH state at $f=\frac{1}{2i+1}$ ($i=1, 2, 3$) has been demonstrated theoretically in the disk geometry for the Laughlin wave function $\Psi_{\rm L}$, which is known to be the exact ground state for certain short-range interactions. The experiments consistently obtain values of $\alpha \!<\! 2i+\!1$, however \cite{Chang01}.

Here we report results of a large numerical study of the microscopic structure of the FQH edge at $f=\frac{1}{3}$. To this end, we diagonalize the interaction Hamiltonian in the disk geometry for up to $N=12$ spin-polarized electrons restricted to lowest Landau level. We construct numerically \cite{preprint} the Laughlin $\Psi_{\rm L}$ and ``Coulomb" $\Psi_{\rm C}$ wave functions as the ground states of the short-range and Coulomb Hamiltonians, respectively, for the total angular momentum $M=\frac{3}{2}N(N-1)$, which gives filling $f=\frac{1}{3}$ in the thermodynamic limit $N\rightarrow \infty$. In particular, we have obtained occupation numbers $\rho (m)$ of the angular momentum basis orbitals

\begin{equation} 
\psi_m(r,\theta)=(2\pi\, 2^m\, m!)^{-1/2}\; r^m\exp (im\theta-r^2/4) ~,
\label{Eq. 1}
\end{equation} 

\noindent     where radius $r$ is in units of magnetic length $\ell =\sqrt {\hbar /eB}$. The Hilbert space is restricted by consideration of orbitals with angular momentum $m\leq m_{\rm max}$ only. For the Haldane $V_1$ short-range interaction, the Laughlin $\rho_{\rm L} (m)$ with $m>m_{\rm max}^{\rm L}=3(N-1)$ vanish identically; for Coulomb interaction good convergence of $\rho_{\rm C} (m)$ is obtained by $m_{\rm max}=m_{\rm max}^{\rm L}+5$ \cite{1/3C}. For example, for $N=12$, the largest $f=\frac{1}{3}$ FQH system studied, $M=198$ and the size of the Hilbert space is 15,293,119 for $m_{\rm max}=35$. Details of this study will be published elsewhere \cite{tobepublished}.

As has been demonstrated by Wen \cite{Wen}, the critical exponent $\alpha$ is equal to the ratio of the occupation numbers 

\begin{equation} 
\alpha=\rho (m_{\rm max}^{\rm L}-1) /\rho (m_{\rm max}^{\rm L} ) 
\label{Eq. 2}
\end{equation} 

\noindent  for the Laughlin state $\Psi_{\rm L}$ on the disk. Wen has also argued that this relationship must hold for any interaction, so long as the FQH state at the same filling $f$ exists, and is not unique for the Laughlin wave function.
In Fig.\ 1 we present the ratio of the occupation numbers for both $\Psi_{\rm L}$ (short-range interaction) and $\Psi_{\rm C}$ (true Coulomb interaction) for $N=3$ to 12. As expected, we obtain $\alpha_{\rm L}=3$ to machine accuracy for $\Psi_{\rm L}$. For Coulomb-interacting  

\begin{figure}
\renewcommand{\baselinestretch}{0.90}
\centerline{\epsfig{file=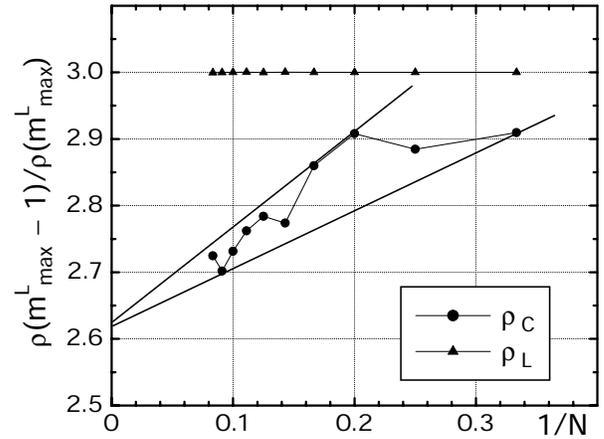, width=7.8cm}}    
\vspace{0.2cm} 
\caption{The ratio of the angular momentum occupation numbers $\rho(m)$ for $N$ interacting electrons on the disk. Shown are the Laughlin state $\rho_{\rm L}$ for a short-range interaction, and the exact ground state for the Coulomb interaction $\rho_{\rm C}$.  }
\label{FIG1}
\end{figure}

\noindent     electrons, the ratio is always less than 3, and an extrapolation to the thermodynamic limit gives $\alpha_{\rm C}=2.62$. While we do not know whether the extrapolation shown in Fig.\ 1 holds for $N>12$, certain other systematic behavior present in the numerical data \cite{tobepublished} allows us to project that in the $N\rightarrow \infty$ limit $2.58\leq \alpha_{\rm C}\leq 2.75$, and is definitely less than 3.

Thus we propose that the deviation of the experimental $\alpha$ from the predicted $\alpha_{\rm L}$ values is not an artefact, and is not due to corrections such as finite bulk diagonal conductivity, disorder, or a variation of the electron density in the sample. Rather, we propose, the effect has a fundamental origin: the value of the critical $I$-$V$ exponent is not universal, but depends on the range of particular interaction.

This work was supported in part by the NSF.

\vspace{0.3cm}
\noindent    V.~J.~Goldman and E.~V.~Tsiper \\
\indent {\small Department of Physics, State University of New York}\\
\indent {\small Stony Brook, NY 11794-3800}\\
January 11, 2001 \\


\bibliographystyle{prsty}

\end{multicols}

\end{document}